\documentclass[12pt,english]{article}
\usepackage[T2A,T2A,T1]{fontenc}
\usepackage[koi8-r]{inputenc}
\usepackage{amsmath}
\usepackage{amssymb}
\usepackage{graphicx}
\usepackage{esint}

\makeatletter
\@ifundefined{date}{}{\date{}}
\AtBeginDocument{\DeclareFontEncoding{T2A}{}{}}\AtBeginDocument{\DeclareFontEncoding{T2A}{}{}}\AtBeginDocument{\DeclareFontEncoding{T2A}{}{}}\AtBeginDocument{\DeclareFontEncoding{T2A}{}{}}\AtBeginDocument{\DeclareFontEncoding{T2A}{}{}}\usepackage{babel}

\newtheorem{theorem}{Theorem}

\newtheorem{lemma}{Lemma}
\newtheorem{remark}{Remark}
\usepackage{babel}

\usepackage{babel}

\makeatother

\usepackage{babel}
\begin{document}

\title{From $N$-body problem to Euler equations}

\author{Lykov A. A., Malyshev V. A. \thanks{Lomonosov Moscow State University, Faculty of Mechanics and Mathematics,
Vorobyevy Gory 1, Moscow, 119991, Russia}}
\maketitle
\begin{abstract}
This paper contains a rigorous mathematical example of direct derivation
of the system of Euler hydrodynamic equations from Hamiltonian equations
for $N$ point particle system as $N\to\infty$. Direct means that
the following standard tools are not used in the proof: stochastic
dynamics, thermodynamics, Boltzmann kinetic equations, correlation
functions approach by N. N. Bogolyubov. 
\end{abstract}

\pagebreak{}

\section{Introduction}

Classical mechanics, from mathematical point of view, is mostly developed
in cases, where two extremely idealized forms of material objects
are assumed - point particles (ordinary differential equations) and
continuum media (partial differential equations). However, big difference
exists in the ideology of these two theories: for point particles
the model is defined by the choice of the interaction potential between
particles, which is supposed to be known, but in the continuum mechanics
the interaction is defined by the pressure, which is one of the unknown
functions in the equations. Many papers - both mission proposals \cite{Bogoliubov,Morrey}
and concrete results for concrete models \cite{March_Pulvirenti,Esposito1,Esposito2,Dobrushin}
- discussed the connections between these two fields. We do not give
here review of these papers as we do not use neither their results
nor methods. Moreover, our approach is direct that is we do not use
any of the following approaches: stochastic dynamics, thermodynamics,
Boltzmann kinetic equations, correlation functions approach by N.
N. Bogolyubov.

Now we make the above claims more precise. Hamiltonian finite particle
system is defined by the system of equations for the particle trajectories
$x_{i}(t)\in R^{d},i=1,...,N$ 
\[
m_{i}\frac{d^{2}x_{i}}{dt^{2}}=-\frac{\partial H}{\partial x_{i}}
\]
with the Hamiltonian $H$.

We define the continuum ($d$-dimensional) media as a bounded open
subset $\Lambda\subset R^{d}$, the dynamics of this media is given
by the system of such domains $\Lambda_{t},t\in[0,T),0<T\leq\infty,$
together with the system of diffeomorphisms $S^{t}:\Lambda=\Lambda_{0}\to\Lambda_{t},t\in[0,T)$,
smooth also in $t$. The trajectory of the point (particle) $x\in\Lambda_{0}$
of the continuum media is the function $y(t,x)=S^{t}x$. The main
unknown variable in the Euler equations is the velocity $u(t,y)$
of the particle, which at time $t$ is at point $y$. This definition
of $u$ has sense iff such particle is unique, that is iff for any
$t$ and any $x_{1}\neq x_{2}$ 
\[
y(t,x_{1})\neq y(t,x_{2})
\]
that is iff the trajectories (particles) $y(t,x)$ do not collide.

This property obviously should be related to the similar property
for $N$ particle system, if we want to obtain continuum media trajectories
in the limit $N\to\infty$ (one could call this the ultralocal limit).

We say that the $N$-particle system has no collisions, if for all
$1\leq j<k\leq N$ and all $t\in[0,\infty)$ 
\[
x_{j}(t)\neq x_{k}(t),
\]
and has strong property of absence of collisions if 
\begin{equation}
\inf_{t\geqslant0}\inf_{j,k:j\neq k}|(x_{j}(t)-x_{k}(t))|>0\label{strong_nonintersection}
\end{equation}
It is evident that there will not be any collisions if the repulsion
between particles is sufficiently strong. However, for general Hamiltonian
systems the following question is completely non trivial: for which
initial conditions $x_{k}(0),\ \dot{x}_{k}(0),k=1,...,N,$ the system
enjoys the absence of collisions property. In this paper the property
(\ref{strong_nonintersection}) plays the central role. It is surprising
that we did not find papers where this property is discussed in the
derivation of continuum media equations. However, it was widely discussed
in Celestial Mechanics (gravitation potential), see for example \cite{Siegel}.

We consider the particle system on the real line with a particular
Lennard-\={J}ones type potential and prove that the particle trajectories
of the $N$-particle system, for $N\to\infty$, converge, in the sense
defined below, to the trajectories of the continuum particle system.
Moreover, we get the system of 3 equations of the Euler type (which
is considered in \cite{ChorinMarsden}) for the functions: $u(t,x)$
- the velocity, $p(t,x)$ - the pressure and $\rho(t,x)$ - the density
\begin{equation}
\rho_{t}+u\rho_{x}+\rho u_{x}=0,\label{mass_conservation_1}
\end{equation}
\begin{equation}
u_{t}+uu_{x}=-\frac{p_{x}}{\rho},\label{euler_1}
\end{equation}
\begin{equation}
p=p(\rho),\label{equation_of_state_general}
\end{equation}
In continuum mechanics these equations correspond to the conservation
laws of mass, momentum and to the thermodynamic equation of state.
In physics the first two equations are quite general. But the third
one depends on the matter type and thermodynamic situation and should
be given separately. In our derivation, all these equations and functions
obtain simple and intuitive mechanical meaning (without probability
theory and thermodynamics) for the $N$-particle system. In particular,
the pressure can be considered as an analog of interaction potential
in Hamiltonian mechanics.

\section{Main Results}

\paragraph{The model}

We consider Hamiltonian system of $N$ particles (of unit mass) with
coordinates $x_{1},...,x_{N}$ on $R$ and the Hamiltonian 
\[
H=\sum_{k=1}^{N}\frac{v_{k}^{2}}{2}+U
\]
The potential energy $U$ of the particle system with the coordinates
$x_{1},x_{2},,...,x_{N}$ is defined by the interaction potential
\[
U=\sum_{1\leq k<l\leq N}\frac{\omega^{2}}{2}I(|x_{k}-x_{l}|),I\in\mathbf{I}(a,a_{1})
\]
where $\mathbf{I}(a,a_{1}),0<a_{1}<a,$ is the class of functions
$I(x)$ on $R_{+}$ with the following two properties

1) $I(x)=(x-a)^{2}$ for $a-a_{1}<x<a+a_{1}$ with some constant $0<a_{1}<a$.

2) $I(x)=const$ for $x\geq a+a_{1}$

3) $I(x)$ is arbitrary for $0<x\leq a-a_{1}$

\paragraph{Scaling}

Our system contains three parameters: $\omega,a$ and $a_{1}$. We
could add also mass but the scaling of mass and/or time could be reduced
to the scaling of $\omega$.

If $N$ is large and all particles are situated on some finite interval
then $a$ should be of order $N^{-1}$. We put $a=\frac{1}{N}$. Then
the system will be in equilibrium (zero force on each particle) iff
for all $k$ $x_{k+1}-x_{k}=\frac{1}{N}$. Correspondingly, we put
$a_{1}=\frac{r}{N}$ for some $0<r<1$ not depending on $N$. The
remaining parameter we choose as 
\begin{equation}
\omega=\omega'N\label{omega}
\end{equation}
for some $\omega'>0$, not depending on $N$,

\paragraph{Initial conditions}

We always assume the following initial conditions 
\begin{equation}
x_{1}(0)=0,\ \dot{x}_{1}(0)=v,\label{initial_1}
\end{equation}
\begin{equation}
x_{k+1}(0)-x_{k}(0)=\frac{1}{N}X\left(\frac{k}{N}\right)>0,\quad\dot{x}_{k+1}(0)-\dot{x}_{k}(0)=\frac{1}{N}V\left(\frac{k}{N}\right),\ k=1,\ldots,N-1\label{initial_other}
\end{equation}
for some $v\in\mathbb{R}$, and for some functions $X,V\in C^{4}([0,1]),$
where $X>0$. Thus, the functions $X$ and $V$ define smooth profile
of the initial conditions. Then the kinetic and potential energies
of the system will be of the order $O(N)$.

It is convenient to assume also that 
\begin{equation}
X(0)=X(1)=1,\ V(0)=V(1)=0,\label{particles_1_N}
\end{equation}
The second condition (\ref{particles_1_N}) means that two leftmost
(two rightmost) particles initially have almost (up to $O(N^{-2})$)
equal velocities, and the first condition (\ref{particles_1_N}) means
that both boundary particles are subjected to almost zero force.

Let $\Omega_{N}=\Omega_{N}(\gamma)$ be the domain of $R^{N}=\{(x_{1},...,x_{N})\}$,
defined for some $0<\gamma<1$ by the estimates 
\[
\frac{1-\gamma}{N}\leqslant x_{k+1}-x_{k}\leqslant\frac{1+\gamma}{N}
\]
uniformly in $k=1,...,N$. We want to prove that if initially our
system is in this region, then under certain conditions it will stay
in $\Omega_{N}$ forever. The obvious corollary is that there will
never be collisions between particles. It might seem that under such
conditions there will not be interesting dynamics, but this is wrong,
see pictures at the end of the paper.

\paragraph{Absence of collisions for $N$-particle case}

The condition below allows to estimate distances between particles
at any time moment. Denote 
\[
\alpha=\int_{0}^{1}|X''(y)|\ dy,\,\,\,\beta=\int_{0}^{1}|V''(y)|\ dy.
\]
Further on we will use the concrete value of $\gamma$, defined in
terms of the main parameters of the system 
\begin{equation}
\gamma=\gamma(\alpha,\beta,\omega')=2\alpha+\frac{\beta N}{\omega}=2\alpha+\frac{\beta}{\omega'}\label{gamma_concrete}
\end{equation}

Further on the condition $\gamma=\gamma(\alpha,\beta,\omega')<\min(r,\frac{1-r}{2})$
is always assumed.

\begin{theorem} \label{Th_bound} Assume that initially our system
is in $\Omega_{N}$. Then it stays in $\Omega_{N}$ forever. That
is for any $t\geqslant0$ and any $k=1,2,\ldots,N-1$ the following
inequalities hold: 
\begin{equation}
\frac{1-\gamma}{N}\leqslant x_{k+1}(t)-x_{k}(t)\leqslant\frac{1+\gamma}{N}\label{gamma_bounds}
\end{equation}
It follows that the particles never collide in the strong sense (\ref{strong_nonintersection}).

\end{theorem}

Note that the scaling of $\omega$ is crucial to create the repulsion
necessary for the particles did not collide.

To understand the importance of the choice of $\alpha$, note that
$X(y)-1$ characterizes the deviation of the chain from the equilibrium,
$X'(y)$ characterizes the speed of change of this equilibrium, and
$\alpha$ can be considered as the full variation. Thus the following
simple statement is useful to estimate such deviation at initial time
moment.

\begin{lemma}\label{X_blizko_1}

For any $y\in[0,1]$ 
\[
1-\alpha\leqslant X(y)\leqslant1+\alpha
\]

\end{lemma}

\paragraph{Strategy of the proof}

If we could prove Theorem \ref{Th_bound} for some potential in the
class $\mathbf{I}(a,a_{1})$, then the $\gamma$-bounds (\ref{gamma_bounds})
indicate that it will also hold for any potential $I\in\mathbf{I}(a,a_{1}).$
In the following proofs we will use the simplest of such potential
- the quadratic potential 
\[
I(x_{k+1}-x_{k})=(x_{k+1}-x_{k}-a)^{2}
\]
Even more, we assume the nearest neighbour interaction for this potential
. Then the following system of linear differential equations holds
\begin{align}
\ddot{x}_{1}= & \ \omega^{2}(x_{2}-x_{1}-a),\label{mainEq1}\\
\ddot{x}_{k}= & \ \omega^{2}(x_{k+1}-x_{k}-a)-\omega^{2}(x_{k}-x_{k-1}-a),\ k=2,3,\ldots,N-1,\label{mainEqk}\\
\ddot{x}_{N}= & \ -\omega^{2}(x_{N}-x_{N-1}-a),\label{mainEqN}
\end{align}
In this case we will also prove $\gamma$-bounds (\ref{gamma_bounds}).

Now from these $\gamma$-bounds we want to show how from this the
Theorem \ref{Th_bound} follows for any class of potentials $\mathbf{I}(\frac{1}{N},\frac{r}{N})$.
It is sufficient to show that for any $t$ and $k$ the following
inequalities hold 
\begin{equation}
\frac{1-r}{N}\leqslant x_{k+1}(t)-x_{k}(t)\leqslant\frac{1+r}{N},\label{r_1}
\end{equation}
\begin{equation}
x_{k+2}(t)-x_{k}(t)>\frac{1+r}{N}\label{r_2}
\end{equation}
Then (\ref{r_1}) obviously holds if $\gamma<r$. From the $\gamma$-bounds
(\ref{gamma_bounds}) we have the estimate 
\[
x_{k+2}(t)-x_{k}(t)\geq2\frac{1-\gamma}{N},
\]
This estimate implies (\ref{r_2}) if $\gamma<\frac{1-r}{2}$.

\paragraph{Convergence to continuous chain dynamics}

Denote $q(t,x)$ the solution of the wave equation 
\[
q_{tt}=(\omega')^{2}q_{xx},
\]
(here and below the lower indices define the derivatives in the corresponding
variables) with fixed boundary conditions 
\begin{equation}
q(t,0)=q(t,1)=0,\label{q_wave_boundary}
\end{equation}
and with the initial conditions: 
\begin{equation}
q(0,x)=X(x)-1,\ q_{t}(0,x)=V(x).\label{q_initial}
\end{equation}

Let $x_{k}^{(N)}(t)$ be the solution of the main system (\ref{mainEq1})-(\ref{mainEqN})
for given $N$. For any fixed $t$ we want to define two functions
(algorithms) which map the set of points $\Lambda_{t}$ of the continuous
media to the set of particles $\{1,2,...,N\}$, that is to the set
of particle coordinates $\{x_{k}^{(N)}(t)\}$ of the $N$-particle
approximation. To do this, we will use two coordinate systems on the
real intervals: $x$ and $z$, where $z(x)$, for $x\in(0,L),L=L(0)=\int_{0}^{1}X(y)dy,$
is uniquely defined from the equation: 
\begin{equation}
\int_{0}^{z(x)}X(x')dx'=x.\label{z_x}
\end{equation}

In the first algorithm to any $z\in(0,1]$ correspond the particle
with number $[zN]$ (the integer part of $zN\in\mathbb{R}$). Note
that for any $0<z\leq1$ there exists $N(z)$ such that $N>N(z)$
we have $1<[zN]\leq N$.

In the second algorithm to any point $x\in[0,L]$ corresponds the
particle with number $k(x,N)=k(x,N,0)$ so that 
\begin{equation}
x_{k(x,N)}^{(N)}(0)\leqslant x<x_{k(x,N)+1}^{(N)}(0)\label{k_x_N_0}
\end{equation}
Due to positivity of $X(y)$ such number is uniquely defined. Then
it is natural to call the function $x_{k(x,N)}^{(N)}(t)$ the $N$-particle
approximation of the trajectory of the particle $x\in[0,L]$ of the
continuum media. By definition we put $x_{N+1}^{(N)}=\infty$.

\begin{theorem} \label{Th_convergence_1} Let the conditions of the
theorem \ref{Th_bound} hold, and assume also (\ref{omega}). Then:

1) for any $0<T<\infty$ uniformly in $t\in[0,T]$ and in $z\in(0,1]$
\begin{equation}
\lim_{N\rightarrow\infty}x_{1}^{(N)}(t)=G(t,0)=vt+(\omega')^{2}\int_{0}^{t}(t-s)q_{x}(s,0)\ ds,\label{G_t_0}
\end{equation}
\begin{equation}
\lim_{N\rightarrow\infty}x_{[zN]}^{(N)}(t)=G(t,z)=G(t,0)+z+\int_{0}^{z}q(t,x')\ dx',\label{G_t_z}
\end{equation}
2) Let $0\leqslant z_{1}<z_{2}\leqslant1$. Then for any $t\geqslant0$
\[
(1-\gamma)(z_{2}-z_{1})\leqslant G(t,z_{2})-G(t,z_{1})\leqslant(1+\gamma)(z_{2}-z_{1}),
\]
where $\gamma$ is defined in (\ref{gamma_concrete}). Otherwise speaking,
the continuum media particles do not collide;

3) for any $T>0$ uniformly in $x\in[0,L]$ and in $t\in[0,T]$ 
\begin{equation}
\lim_{N\rightarrow\infty}x_{k(x,N)}^{(N)}(t)=y(t,x)=G(t,z(x)),\label{G_t_z_x}
\end{equation}
4) the function $G(t,z)$ satisfies the wave equation 
\begin{equation}
\frac{d^{2}G(t,z)}{dt^{2}}=(\omega')^{2}\frac{d^{2}G(t,z)}{dz^{2}}\label{G_wave}
\end{equation}
with boundary conditions $G_{z}(t,0)=G_{z}(t,1)=1$ and initial conditions
\[
G(0,z)=\int_{0}^{z}X(x')dx',\quad G_{t}(0,z)=v+\int_{0}^{z}V(x')dx'.
\]
\end{theorem}

An obvious corollary is that for any $T>0$, uniformly in $t\in[0,T]$,
we have the following asymptotic limit for the length $L_{N}(t)=x_{N}(t)-x_{1}(t)$
of the chain 
\[
\lim_{N\rightarrow\infty}L_{N}(t)=L(t)=1+\int_{0}^{1}q(t,x')dx'
\]

\paragraph{Continuity equation (mass conservation law)}

Further on, the function $y(t,x)$ will be called the trajectory of
the particle $x\in[0,L]$. Then the particles do not collide and one
can unambiguously define the function $u(t,y)$ as the speed of the
(unique) particle situated at time $t$ at the point $y$ , that is
\begin{equation}
u(t,y(t,x))=\frac{dy(t,x)}{dt}.\label{yveq}
\end{equation}
Also we will need the notation: 
\[
Y_{0}(t)=y(t,0),\quad Y_{L}(t)=y(t,L).
\]
For given $N$ we define the distribution function at time $t$: 
\[
F^{(N)}(t,y)=\frac{1}{N}|\{k\in\{1,2,\ldots,N\}:\ x_{k}^{(N)}(t)\leqslant y\}|,\ y\in\mathbb{R}
\]
where $|\cdot|$ is the number of particle in the set.

\begin{lemma} \label{Lemma_distrFunc} Denote $x(t,y)\in[0,L]$ the
(unique) particle, which reached the point $y$ at time $t$, that
is 
\begin{equation}
y(t,x(t,y))=y.\label{xtydef}
\end{equation}
Then uniformly in $y\in[Y_{0}(t),Y_{L}(t)]$ and in $t\in[0,T]$,
for any $T<\infty$, we have 
\[
\lim_{N\rightarrow\infty}F^{(N)}(t,y)=z(x(t,y))=F(t,y),
\]
where the (smooth) function $z(x)$ is the same as the one introduced
in (\ref{z_x}).

\end{lemma}

In connection with Lemma \ref{Lemma_distrFunc} define the density
by the formula 
\begin{equation}
\rho(t,y)=\frac{dF(t,y)}{dy}=\frac{dz(x(t,y))}{dy}\label{density}
\end{equation}

\begin{theorem}\label{Th_mass_conservation} For any $t\geqslant0,\ y\in[Y_{0}(t),Y_{L}(t)]$
\begin{equation}
\frac{\partial\rho(t,y)}{\partial t}+\frac{d}{dy}\left(u(t,y)\rho(t,y)\right)=0.\label{mass conservation_2}
\end{equation}
\end{theorem}

\paragraph{Euler equation (momentum conservation) and equation of state}

\begin{theorem} \label{Th_Euler} For any $t\geqslant0,\ y\in[Y_{0}(t),Y_{L}(t)]$
we have: 
\begin{equation}
\frac{\partial u(t,y)}{\partial t}+u(t,y)\frac{\partial u(t,y)}{\partial y}=-(\omega')^{2}\frac{1}{\rho(t,y)}\frac{d}{dy}\frac{1}{\rho(t,y)}=-\frac{1}{\rho(t,y)}\frac{dp(t,y)}{dy},\label{EulerEq_2}
\end{equation}
if we put 
\begin{equation}
p(t,y)=-\frac{(\omega')^{2}}{\rho(t,y)}+C.\label{state_equation_2}
\end{equation}
\end{theorem}

Constant $C$ can be chosen as 
\[
C=(\omega')^{2},
\]
so that at equilibrium (when $\rho=1$) the pressure were zero.

\subsubsection*{Right side of the Euler equation as the limit of interaction forces}

For given $y$ and $t$ define the number $k(y,N,t)$ so that 
\begin{equation}
x_{k(y,N,t)}^{(N)}(t)\leqslant y<x_{k(y,N,t)+1}^{(N)}(t)\label{k_y_N_t}
\end{equation}
Consider the point $y\in[Y_{0}(t),Y_{L}(t)]$ and the force acting
on the particle with number $k(y,N,t)$: 
\[
R^{(N)}(t,y)=\omega^{2}\left(x_{k(y,N,t)+1}^{(N)}-x_{k(y,N,t)}^{(N)}-\frac{1}{N}\right)-\omega^{2}\left(x_{k(y,N,t)}^{(N)}-x_{k(y,N,t)-1}^{(N)}-\frac{1}{N}\right)
\]
\begin{theorem}\label{Th_force} Let the conditions of the theorem
\ref{Th_convergence_1} hold. Then for any $0<T<\infty$, uniformly
in $y\in[Y_{0}(t),Y_{L}(t)]$ and in $t\in[0,T]$ the following equality
holds: 
\[
\lim_{N\rightarrow\infty}R^{(N)}(t,y)=R(t,y)=-\frac{1}{\rho(t,y)}\frac{dp(y)}{dy},
\]
where the functions $p,\rho$ are the same as in theorem \ref{Th_Euler}.

\end{theorem}

Thus, the pressure can be considered as a continuous interaction potential
for continuum media, an analog of interaction potentials in Hamiltonian
particle mechanics.

\paragraph{Limit of the energy}

Define the potential and kinetic energy of the particle with number
$k(y,N,t)$ at time $t$ for $N$-particle approximation correspondingly
as: 
\begin{align*}
U^{(N)}(t,y)= & \frac{1}{4}\omega^{2}\left(x_{k(y,N,t)+1}^{(N)}-x_{k(y,N,t)}^{(N)}-\frac{1}{N}\right)^{2}+\frac{1}{4}\omega^{2}\left(x_{k(y,N,t)}^{(N)}-x_{k(y,N,t)-1}^{(N)}-\frac{1}{N}\right)^{2},\\
T^{(N)}(t,y)= & \ \frac{1}{2}\left(\dot{x}_{k(y,N,t)}^{(N)}\right)^{2}.
\end{align*}
\begin{theorem}\label{Th_Energy} For any $t,y$ (uniformly as in
the previous theorem) the following limits hold: 
\begin{align*}
U(t,y)=\lim_{N\rightarrow\infty}U^{(N)}(t,y)= & \ \frac{1}{2}(\omega')^{2}\left(\frac{1}{\rho(t,y)}-1\right)^{2}=\frac{1}{2(\omega')^{2}}p^{2}(t,y),\\
T(t,y)=\lim_{N\rightarrow\infty}T^{(N)}(t,y)= & \ \frac{1}{2}u^{2}(t,y)
\end{align*}

\end{theorem}

\section{Proofs}

As we explained previously, we shall always use the system (\ref{mainEq1})-(\ref{mainEqN}).

\subsection{Proof of Theorem \ref{Th_bound}}

\paragraph{Proof of Lemma \ref{X_blizko_1}}

Put $f(y)=X(y)-1$ and use the following Lemma \ref{zeroEndsLemma}.

\begin{lemma} \label{zeroEndsLemma} Assume that $f\in C^{2}([0,1])$
and $f(0)=f(1)=0$. Then the following inequality holds: 
\[
\sup_{y\in[0,1]}|f(y)|\leqslant\int_{0}^{1}|f''(y)|\ dy.
\]
\end{lemma} In fact, 
\[
f(y)=\int_{0}^{y}f'(x)\ dx.
\]
It follows that 
\[
\sup_{y\in[0,1]}|f(y)|\leqslant\int_{0}^{1}|f'(x)|\ dx.
\]
As $f(0)=f(1)=0$, then there exists point $x^{*}\in(0,1)$ such that
\[
f'(x^{*})=0.
\]
Thus we have: 
\[
\int_{0}^{1}|f'(x)|\ dx=\int_{0}^{1}\left|\int_{x}^{x^{*}}f''(u)du\right|\ dx\leqslant\sup_{x\in[0,1]}\left|\int_{x}^{x^{*}}f''(u)du\right|\leqslant\int_{0}^{1}|f''(u)|\ du.
\]
This proves the Lemma.

\begin{remark}

The set of functions $\Sigma$, satisfying the conditions of Lemma
\ref{zeroEndsLemma} is a linear space. Moreover, 
\[
\alpha(f)=\int_{0}^{1}|f''(y)|\ dy.
\]
defines a norm on this space. Lemma \ref{zeroEndsLemma} states that
the uniform norm does not exceed the norm $\alpha(\cdot)$. However,
we want to note that these two norms are not equivalent. In fact,
assume the contrary, i. e. that there exists constant $a>0$ such
that for any function $f\in\Sigma$ 
\[
\alpha(f)\leqslant a\sup_{y\in[0,1]}|f(y)|.
\]
Then put $f_{k}(y)=\sin\pi ky$. Then $\alpha(f_{k})\rightarrow\infty$
as $k\rightarrow\infty$, but $\sup_{y\in[0,1]}|f_{k}(y)|=1$. This
is a contradiction.

\end{remark}

\paragraph{Deviation variables}

Define the deviation variables $q_{k}(t)=x_{k+1}(t)-x_{k}(t)-a,\ k=1,\ldots,N-1$,
and put by definition $q_{0}=q_{N}=0$, Then the functions $q_{k}$
satisfy the equations: 
\begin{equation}
\ddot{q}_{k}=\omega^{2}(q_{k+1}-q_{k})-\omega^{2}(q_{k}-q_{k-1})=\omega^{2}(q_{k+1}-2q_{k}+q_{k-1}),\ k=1,2,\ldots,N-1\label{diffeqk}
\end{equation}
with initial conditions which follow from (\ref{initial_other}) 
\[
q_{k}(0)=\frac{1}{N}X\left(\frac{k}{N}\right)-\frac{1}{N},\quad\dot{q}_{k}(0)=\frac{1}{N}V\left(\frac{k}{N}\right),\ k=0,1,\ldots,N
\]
In fact, from equations (\ref{mainEq1})-(\ref{mainEqN}) we have
\[
\ddot{x}_{k}=\omega^{2}(q_{k}-q_{k-1}),\,\,\,k=1,...,N,
\]
\[
\ddot{x}_{k}=\ddot{x}_{k+1}-\ddot{q}_{k},k=1,...,N-1,
\]
Then for $k=1,\ldots,N-1$ 
\[
\omega^{2}(q_{k+1}-q_{k})-\ddot{q}_{k}=\omega^{2}(q_{k}-q_{k-1}).
\]
The last equality is equivalent to (\ref{diffeqk}).

\begin{remark}

The inverse transformation is given by 
\begin{equation}
x_{k}(t)=x_{1}(t)+\frac{k-1}{N}+\sum_{i=1}^{k-1}q_{i}(t),\quad k=2,3,\ldots,N.\label{xkqk}
\end{equation}
and for $x_{1}(t)$, by definition of $q_{1}$, we have the equation:
\[
\ddot{x}_{1}=\omega^{2}q_{1}.
\]
It follows 
\begin{equation}
x_{1}(t)=x_{1}(0)+\dot{x}_{1}(0)t+\omega^{2}\int_{0}^{t}ds\int_{0}^{s}q_{1}(s')\ ds'=x_{1}(0)+\dot{x}_{1}(0)t+\omega^{2}\int_{0}^{t}(t-s)q_{1}(s)\ ds\label{x1formula}
\end{equation}
The last equality follows from comparison of derivatives of both sides.

\end{remark}

One can rewrite the system (\ref{diffeqk}) in the matrix form: 
\[
\ddot{q}=-Wq,
\]
where the matrix $W$ is a three diagonal non negative definite $(n\times n)$-matrix
with $n=N-1$, and $q=(q_{1}(t),\ldots,q_{n}(t))^{T}$ is a column
vector.

\paragraph{Spectrum of the matrix $W$ }

We will show that $W$ is positive definite, and will find the basis
$v_{1},\ldots,v_{n}$ of eigenvectors of $W$, with corresponding
eigenvalues $\lambda_{1},\ldots,\lambda_{n}$: $Wv_{j}=\lambda_{j}v_{j},\ j=1,\ldots,n$.

Let $e_{k}$ be the standard unit coordinate vectors in $\mathbb{R}^{n}$.
For $j=1,\ldots,N-1$ define vectors $v_{j}$ by 
\begin{equation}
y_{j}(k)=(v_{j},e_{k})=\sqrt{\frac{2}{N}}\sin\frac{\pi jk}{N},\ k=1,2,\ldots,N-1.\label{specDef}
\end{equation}
and the numbers $\lambda_{j}$ by 
\[
\lambda_{j}=4\omega^{2}\sin^{2}\frac{\pi j}{2N}.
\]
Let us prove that $v_{j}$ are eigenvectors of $W$ with eigenvalues
$\lambda_{j}$. Note that if we define $y_{j}(k)$ from (\ref{specDef})
also at the points $k=0$ and $k=N$, then we get $y_{j}(0)=y_{j}(N)=0$.
Then for all $k=1,\ldots,N-1$ 
\[
(Vv_{j},e_{k})=-\omega^{2}(y_{j}(k+1)-y_{j}(k))+\omega^{2}(y_{j}(k)-y_{j}(k-1))=
\]
\[
=2\omega^{2}\sqrt{\frac{2}{N}}\left(-\sin\frac{\pi j}{2N}\cos\frac{\pi j(2k+1)}{2N}+\sin\frac{\pi j}{2N}\cos\frac{\pi j(2k-1)}{2N}\right)
\]
\[
=-2\omega^{2}\sqrt{\frac{2}{N}}\sin\frac{\pi j}{2N}\left(\cos\frac{\pi j(2k+1)}{2N}-\cos\frac{\pi j(2k-1)}{2N}\right)=
\]
\[
=4\omega^{2}\sqrt{\frac{2}{N}}\sin\frac{\pi j}{2N}\sin\frac{\pi jk}{N}\sin\frac{\pi j}{2N}=\lambda_{j}y_{j}(k).
\]
where $(,)$ is the standard scalar product in $\mathbb{R}^{n}$.
As all $\lambda_{j}$ positive and different, then $W$ is positive
definite.

\paragraph{Dynamics of deviations}

\begin{lemma} Let for any $j=1,\ldots,N-1$ 
\[
Q_{j}=\sqrt{\frac{2}{N}}\sum_{i=1}^{N-1}q_{i}(0)\sin\frac{\pi ij}{N},\quad P_{j}=\sqrt{\frac{2}{N}}\sum_{i=1}^{N-1}\dot{q}_{i}(0)\sin\frac{\pi ij}{N},\quad\omega_{j}=2\omega\sin\frac{\pi j}{2N}.
\]
Then 
\[
q_{k}(t)=\sqrt{\frac{2}{N}}\sum_{j=1}^{N-1}\left(Q_{j}\cos\omega_{j}t+P_{j}\frac{\sin\omega_{j}t}{\omega_{j}}\right)\sin\frac{\pi jk}{N},
\]
\end{lemma}

Proof. Using the expansion of $q(t)$ in the basis 
\[
q(t)=\sum_{j=1}^{n}Q_{j}(t)v_{j},\quad Q_{j}(t)=(q(t),v_{j}),
\]
we will get equations for $Q_{j}(t)$: 
\[
\ddot{Q}_{j}=-\lambda_{j}Q_{j},
\]
This gives 
\[
Q_{j}(t)=Q_{j}(0)\cos\omega_{j}t+\dot{Q}_{j}(0)\frac{\sin\omega_{j}t}{\omega_{j}},\quad\omega_{j}=\sqrt{\lambda_{j}}.
\]
Then 
\[
q_{k}(t)=(q(t),e_{k})=\sum_{j=1}^{n}Q_{j}(t)(v_{j},e_{k})=\sum_{j=1}^{n}\left(Q_{j}(0)\cos\omega_{j}t+\dot{Q}_{j}(0)\frac{\sin\omega_{j}t}{\omega_{j}}\right)(v_{j},e_{k}),
\]
and the Lemma is proved.

\paragraph{Estimate of the coefficients $Q_{j},P_{j}$ }

We have 
\[
Q_{j}=\sqrt{\frac{2}{N}}\sum_{i=1}^{N-1}q_{i}(0)\sin\frac{\pi ij}{N}=(q(0),v_{j})=\frac{1}{\lambda_{j}}(q(0),Vv_{j})=\frac{1}{\lambda_{j}}(Vq(0),v_{j})=
\]
\[
=-\frac{\omega^{2}}{\lambda_{j}}\sqrt{\frac{2}{N}}\sum_{i=1}^{N-1}\left((q_{i+1}(0)-q_{i}(0))-(q_{i}(0)-q_{i-1}(0))\right)\sin\frac{\pi ij}{N}.
\]
Then let us estimate the sum 
\[
S_{N}=\sum_{i=1}^{N-1}\left((q_{i+1}(0)-q_{i}(0))-(q_{i}(0)-q_{i-1}(0))\right)\sin\frac{\pi ij}{N}
\]
where 
\[
q_{i}(0)=\frac{1}{N}X(\frac{i}{N})-\frac{1}{N},\,\,\,i=0,1,...,N
\]
We have 
\[
q_{i+1}(0)-q_{i}(0)=\frac{1}{N}\left(X\left(\frac{i+1}{N}\right)-X\left(\frac{i}{N}\right)\right)=\frac{1}{N^{2}}X'(\theta_{i}),
\]
for some point $\theta_{i}\in(\frac{i}{N},\frac{i+1}{N})$. This gives
\[
|S_{N}|=\frac{1}{N^{2}}\left|\sum_{i=1}^{N-1}(X'(\theta_{i})-X'(\theta_{i-1}))\sin\frac{\pi ij}{N}\right|\leqslant\frac{1}{N^{2}}\sum_{i=1}^{N-1}\left|X'(\theta_{i})-X'(\theta_{i-1}))\right|\leqslant\frac{1}{N^{2}}\int_{0}^{1}|X''(y)|\ dy.
\]
and thus 
\[
|Q_{j}|\leqslant\frac{1}{4\sin^{2}\frac{\pi j}{2N}}\sqrt{\frac{2}{N}}\frac{\alpha}{N^{2}},\quad\alpha=\int_{0}^{1}|X''(y)|\ dy.
\]
Taking into account the inequality $\omega_{j}=\omega\sin\frac{\pi j}{2N}\geqslant\omega\frac{j}{N}$,
we get 
\[
|Q_{j}|\leqslant\frac{1}{4}\sqrt{\frac{2}{N}}\frac{\alpha}{j^{2}}.
\]
Similar estimates holds for $P_{j}$: 
\[
|P_{j}|\leqslant\frac{1}{4}\sqrt{\frac{2}{N}}\frac{\beta}{j^{2}},\quad\beta=\int_{0}^{1}|V''(y)|\ dy.
\]
This gives the final estimate 
\[
|q_{k}(t)|\leqslant\frac{1}{4}\frac{2}{N}\sum_{k=1}^{N-1}(\frac{\alpha}{j^{2}}+\frac{\beta N}{2\omega j^{3}})\leqslant\frac{\gamma}{N},\quad\gamma=2\alpha+\frac{\beta N}{\omega}.
\]

\subsection{Proof of Theorem \ref{Th_convergence_1}}

We will denote now $q_{k}(t)=q_{k}^{(N)}(t)$, emphasizing the dependence
on $N$.

\begin{lemma} \label{qklimit} Assume the conditions of Theorem \ref{Th_convergence_1}.
Then for any $T>0$ 
\[
\max_{t\in[0,T]}\max_{k=1,\ldots,N-1}\left|q_{k}^{(N)}(t)-\frac{1}{N}q(t,\frac{k}{N})\right|\leqslant\frac{c\ln N}{N^{3}},
\]
for some constant $c>0$ not depending on $N$. \end{lemma}

Proof. Consider the difference 
\[
\Delta_{k}^{(N)}(t)=q_{k}^{(N)}(t)-\frac{1}{N}q(t,\frac{k}{N}),\ k=0,\ldots,N
\]
For any $k=1,\ldots,N-1$ we have 
\[
\ddot{\Delta}_{k}^{(N)}(t)=\omega^{2}(q_{k+1}^{(N)}-q_{k}^{(N)})-\omega^{2}(q_{k}^{(N)}-q_{k-1}^{(N)})-\frac{1}{N}(\omega')^{2}q_{xx}(t,\frac{k}{N})
\]
Note that for all $k=1,\ldots,N-1$ 
\[
(q(t,\frac{k+1}{N})-q(t,\frac{k}{N}))-(q(t,\frac{k}{N})-q(t,\frac{k-1}{N}))=\frac{1}{N^{2}}q_{xx}(t,\frac{k}{N})+r_{k}^{(N)}(t),
\]
and moreover the remainder term can be estimated as 
\[
|r_{k}^{(N)}(t)|\leqslant\frac{1}{12N^{4}}\max_{t\in[0,T]}\max_{x\in[0,1]}\left|\frac{d^{4}q(t,x)}{dx^{4}}\right|=\frac{c_{1}}{N^{4}}.
\]
Then we have the equations 
\[
\ddot{\Delta}_{k}^{(N)}(t)=\omega^{2}(\Delta_{k+1}^{(N)}-\Delta_{k}^{(N)})-\omega^{2}(\Delta_{k}^{(N)}-\Delta_{k-1}^{(N)})+r_{k}^{(N)}(t)(\omega')^{2}N
\]
with initial conditions 
\[
\Delta_{k}^{(N)}(0)=0,\quad\dot{\Delta}_{k}^{(N)}(0)=0,\,\,\,k=0,...,N
\]
Introduce the vectors 
\[
\Delta^{(N)}(t)=(\Delta_{1}^{(N)}(t),\ldots,\Delta_{N-1}^{(N)}(t))^{T},\quad r^{(N)}(t)=(r_{1}^{(N)}(t),\ldots,r_{N-1}^{(N)}(t))^{T}.
\]
Then we have the equation 
\[
\ddot{\Delta}^{(N)}=-W\Delta^{(N)}+r^{(N)}(t)(\omega')^{2}N,
\]
where the matrix $W$ was introduced in the proof of Theorem \ref{Th_bound}.
It is easy to see that the solution of this equation is 
\[
\Delta^{(N)}(t)=(\omega')^{2}N\int_{0}^{t}(\sqrt{W})^{-1}\sin\sqrt{W}(t-s)r^{(N)}(s)\ ds,
\]
where $\sqrt{W}$ is the positive definite square root of the matrix
$W$. Thus 
\[
\Delta_{k}^{(N)}(t)=(\Delta^{(N)}(t),e_{k})=(\omega')^{2}N\sum_{j=1}^{N-1}\frac{(v_{j},e_{k})}{\omega_{j}}\int_{0}^{t}\sin\omega_{j}(t-s)(r^{(N)}(s),v_{j})\ ds.
\]
For all $s\in[0,t],\ j=1,\ldots,N-1$ we have the inequality: 
\[
|(r^{(N)}(s),v_{j})|\leqslant\sqrt{\frac{2}{N}}\frac{c_{1}}{N^{3}}
\]
The consequence is that for all $t\in[0,T],\ k=1,\ldots,N-1$ the
following estimate holds: 
\[
|\Delta_{k}^{(N)}(t)|\leqslant(\omega')^{2}N\frac{2}{N}\frac{c_{1}}{N^{3}}\sum_{j=1}^{N-1}\frac{T}{2N\omega'\sin\frac{\pi j}{2N}}\leqslant(\omega')^{2}N\frac{1}{N}\frac{Tc_{1}}{\omega'N^{3}}\sum_{j=1}^{N-1}\frac{1}{j}\leqslant\frac{c_{2}\ln N}{N^{3}},
\]
for some constant $c_{2}>0$, not depending on $N$. The Lemma is
proved.

\paragraph{Proof of the assertion 1) of Theorem \ref{Th_convergence_1} }

Note that for any $t\in[0,T]$ 
\[
q(t,\frac{1}{N})=\frac{1}{N}q_{x}(t,0)+\frac{r(t)}{N^{2}},
\]
where $|r(t)|\leqslant c$ for some constant $c>0$, not depending
on $N$. That is why from the equality (\ref{x1formula}) and Lemma
\ref{qklimit}, we get that uniformly in $t\in[0,T]$ the following
limiting equality holds 
\[
\lim_{N\rightarrow\infty}x_{1}^{(N)}(t)=vt+(\omega')^{2}\int_{0}^{t}(t-s)q_{x}(s,0)\ ds.
\]
Using the equality (\ref{xkqk}) and Lemma \ref{qklimit}, we get:
\[
x_{[zN]}^{(N)}(t)=x_{1}(t)+z+\frac{1}{N}\sum_{k=1}^{[zN]-1}q(t,\frac{k}{N})+r^{(N)}(t,z),
\]
and moreover, there exists constant $C>0$ such that $|r^{(N)}(t,z)|\leqslant\frac{C}{N}$
for all $z\in[0,1],\ t\in[0,T]$. Taking the limit in this equality
we get the assertion of the Theorem.

\paragraph{Proof of assertion 2) }

From evident equality 
\[
x_{[z_{2}N]}(t)-x_{[z_{1}N]}(t)=\sum_{k=[z_{1}N]}^{[z_{2}N]-1}(x_{k+1}(t)-x_{k}(t))
\]
and from Theorem \ref{Th_bound} we get the estimate: 
\[
\frac{1-\gamma}{N}([z_{2}N]-[z_{1}N])\leqslant x_{[z_{2}N]}(t)-x_{[z_{1}N]}(t)\leqslant\frac{1+\gamma}{N}([z_{2}N]-[z_{1}N]).
\]
Taking the limit here we get the assertion.

\paragraph{Proof of assertion 3)}

Firstly, let us prove that for some constant $c>0$, not depending
on $N$, for all $x\in[0,L]$ 
\begin{equation}
\left|\frac{k(x,N)}{N}-z(x)\right|\leqslant\frac{c}{N}\label{k_x_N_inequality}
\end{equation}
Denote 
\[
f(z)=\int_{0}^{z}X(x')dx'.
\]
Then we have $f(z(x))=x$. On the other side, the integral can be
calculated as follows 
\[
f(\frac{k(x,N)}{N})=\frac{1}{N}\sum_{i=1}^{k(x,N)}X(\frac{i}{N})+r_{N}(x)=x_{k(x,N)+1}(0)+r_{N}(x),
\]
where the remainder term enjoys the following estimate: 
\[
|r_{N}(x)|\leqslant\frac{1}{N}\max_{y\in[0,1]}|X'(y)|=\frac{c_{1}}{N}.
\]
By definition of $k(x,N)$ we have: 
\[
x-\frac{c_{1}}{N}\leqslant f(\frac{k(x,N)}{N})<x+\frac{1}{N}X(\frac{k(x,N)}{N})+\frac{c_{1}}{N}<x+\frac{c_{2}}{N},\ c_{2}=c_{1}+\max_{y\in[0,1]}X(y).
\]
The following inequality follows: 
\[
|f(\frac{k(x,N)}{N})-f(z(x))|\leqslant\frac{c_{2}}{N}.
\]
But also for some point $\theta\in[0,1]$ 
\[
f(\frac{k(x,N)}{N})-f(z(x))=(\frac{k(x,N)}{N}-z(x))f'(\theta),
\]
This gives 
\[
\left|\frac{k(x,N)}{N}-z(x)\right|\leqslant\frac{1}{\min_{y\in[0,1]}X(y)}\frac{c_{2}}{N}=\frac{c}{N}.
\]
From the proved inequality (\ref{k_x_N_inequality}) it follows that
\begin{equation}
|k(x,N)-[z(x)N]|\leqslant c',\ c'=c+1\label{k_N_bound_1}
\end{equation}
Then by Theorem \ref{Th_bound} 
\[
|x_{[z(x)N]}(t)-x_{k(x,N)}(t)|\leqslant c\frac{1+\gamma}{N}
\]
Taking the limit in the last inequality we get the assertion.

\paragraph{Proof of Lemma 2}

We will use the particle numbers $k(y,N,t)$, introduced in (\ref{k_y_N_t}).
By definition we take $x_{0}^{(N)}(t)=Y_{0}(t),\ x_{N+1}^{(N)}(t)=Y_{L}(T)$.
It is clear that 
\[
F^{(N)}(t,y)=\frac{k(y,N,t)}{N}.
\]
Further on for given $N$ we consider particle trajectories for the
initial points $x_{k(y,N,t)}^{(N)}(0)$ and $x_{k(x(t,y),N)}^{(N)}(0)$.
We want to prove that at time $t$ the distance between them does
not exceed $c/N$. Using theorem \ref{Th_bound}, we will show that
$k(y,N,t)$ differs from $k(x(t,y),N)$ not more than on some constant.
Lemma will follow from this. Now we give the formal proof. We use
the inequalities: 
\[
|x_{k(x(t,y),N)}^{(N)}(t)-x_{k(y,N,t)}^{(N)}(t)|\leqslant|x_{k(x(t,y),N)}^{(N)}(t)-y(t,x(t,y))|+|x_{k(y,N,t)}^{(N)}(t)-y|
\]
By assertions 1), 2), 3) of Theorem 2, and its proof, we can conclude,
that the following inequality holds: 
\[
|x_{k(x(t,y),N)}^{(N)}(t)-y(t,x(t,y)))|\leqslant\frac{c_{1}}{N},
\]
for some constant $c_{1}>0$ not depending on $N$ and $y$. Then
by definition of $k(y,N,t)$ and Theorem \ref{Th_bound} we have the
estimate for $0<k(y,N,t)<N$: 
\[
|x_{k(y,N,t)}^{(N)}(t)-y|\leqslant|x_{k(y,N,t)}^{(N)}(t)-x_{k(y,N,t)+1}^{(N)}(t)|\leqslant\frac{c_{2}}{N},
\]
for some constant $c_{2}>0$ not depending on $N,y$, . In cases $k(y,N,t)=N$
and $k(y,N,t)=0$ the latter inequality follows from Theorem \ref{Th_convergence_1}.
Then 
\[
|x_{k(x(t,y),N)}^{(N)}(t)-x_{k(y,N,t)}^{(N)}(t)|\leqslant\frac{c}{N},\quad c=c_{1}+c_{2}.
\]
From this inequality and Theorem \ref{Th_bound} we have 
\begin{equation}
|k(x(t,y),N)-k(y,N,t)|\leqslant c',\label{k_N_bound_2}
\end{equation}
for some constant $c'>0$, not depending on $N,y$. We can conclude
that 
\[
\lim_{N\rightarrow\infty}\frac{k(y,N,t)}{N}=\lim_{N\rightarrow\infty}\frac{k(x(t,y),N)}{N}=z(x(t,y)),
\]
where the latter equality follows from the proof of Theorem \ref{Lemma_distrFunc},
assertion 3. The Lemma is thus proved.

\paragraph{Proof of assertion 4)}

The simple calculation gives with (\ref{q_wave_boundary}),(\ref{G_t_z})
\[
\frac{d^{2}G(t,z)}{dz^{2}}=q_{z}(t,z).
\]
\[
\frac{d^{2}G(t,z)}{dt^{2}}=\frac{d^{2}G(t,0)}{dt^{2}}+\int_{0}^{z}q_{tt}(t,z')dz'=(\omega')^{2}q_{z}(t,0)+(\omega')^{2}\int_{0}^{z}q_{zz}(t,z')dz'=
\]
\[
=(\omega')^{2}(q_{z}(t,z)-q_{z}(t,0))+(\omega')^{2}q_{z}(t,0)=(\omega')^{2}q_{z}(t,z)
\]
The boundary and initial conditions can be easily found from the corresponding
conditions on the function $q(t,x)$

\subsection{Proof of Theorem \ref{Th_mass_conservation}}

By definition (\ref{density}) we have 
\[
\rho(t,y)=z'(x(t,y))x_{y}(t,y).
\]
On the other side, differentiation in $y$ of the equality (\ref{xtydef})
gives: 
\begin{equation}
x_{y}(t,y)=\frac{1}{y_{x}(t,x(t,y))}.\label{xderivy}
\end{equation}
Hence, 
\begin{equation}
y_{x}(t,x(t,y))=\frac{z'(x(t,y))}{\rho(t,y)}.\label{yderivx}
\end{equation}
By definition 
\[
\frac{\partial\rho(t,y)}{\partial t}=\frac{d}{dy}\frac{dz(x(t,y)}{dt}=\frac{d}{dy}\left(x_{t}(t,y)z'(x(t,y))\right).
\]
Differentiating in $t$ the equality (\ref{xtydef}) we get: 
\[
x_{t}(t,y)=-\frac{\frac{\partial y(t,x(t,y))}{\partial t}}{y_{x}(t,x(t,y))}=-\frac{u(t,y)}{y_{x}(t,x(t,y))}=-\frac{u(t,y)\rho(t,y)}{z'(x(t,y))}.
\]
The theorem is thus proved.

\subsection{Proof of Theorem \ref{Th_Euler}}

We need the following Lemma.

\begin{lemma} For all $t\geqslant0,\ x\in[0,L]$ 
\[
\frac{\partial u(t,y(t,x))}{\partial t}+u(t,y(t,x))\frac{\partial u(t,y(t,x))}{\partial y}=
\]
\begin{equation}
=(\omega')^{2}G_{zz}(t,z(x))=(\omega')^{2}\frac{y_{xx}(t,x)-\frac{z''(x)}{z'(x)}y_{x}(t,x)}{[z'(x)]^{2}}.\label{velocityEqLemma}
\end{equation}
\end{lemma}

Proof of the Lemma. The left hand side of the formula (\ref{velocityEqLemma})
is the complete derivative of $u(t,y(t,x))$ in $t$, that is $\frac{du(t,y(t,x))}{dt}$.
On the other side, we have by definition: 
\[
\frac{du(t,y(t,x))}{dt}=\frac{d^{2}y(t,x)}{dt^{2}}=\frac{d^{2}G(t,z(x))}{dt^{2}}=(\omega')^{2}\frac{d^{2}G(t,z(x))}{dz^{2}}.
\]
Moreover\c{ } the following formulas hold: 
\begin{align}
y_{x}(t,x)= & \ \frac{dy(t,x)}{dx}=\frac{dG(t,z(x))}{dx}=z'(x)\frac{dG(t,z(x))}{dz}=z'(x)G_{z}(t,z(x)),\label{y_po_x}\\
y_{xx}(t,x)= & \ \frac{d^{2}y(t,x)}{dx^{2}}=[z'(x)]^{2}\frac{d^{2}G(t,z(x))}{dz^{2}}+z''(x)\frac{dG(t,z(x))}{dz}=[z'(x)]^{2}G_{zz}(t,z(x))+z''(x)G_{z}(t,z(x)).\nonumber 
\end{align}
Then 
\[
G_{zz}(t,z(x))=\frac{y_{xx}(t,x)-\frac{z''(x)}{z'(x)}y_{x}(t,x)}{[z'(x)]^{2}}.
\]
and the Lemma is proved.

Let us prove now theorem \ref{Th_Euler}. Putting $x=x(t,y)$ in the
equation (\ref{velocityEqLemma}) gives the following equation: 
\[
\frac{\partial u(t,y)}{\partial t}+u(t,y)\frac{\partial u(t,y)}{\partial y}=R(t,y),
\]
where we introduced the function: 
\begin{equation}
R(t,y)=(\omega')^{2}\frac{y_{xx}(t,x(t,y))-\frac{z''(x(t,y))}{z'(x(t,y))}y_{x}(t,x(t,y))}{[z'(x(t,y))]^{2}}.\label{R_t_y}
\end{equation}
Differentiating the equality (\ref{yderivx}) in $y$ and using (\ref{xderivy}),
we get: 
\[
y_{xx}(t,x(t,y))=\frac{1}{x_{y}(t,y)}\frac{d}{dy}\frac{z'(x(t,y))}{\rho(t,y)}=\frac{z'(x(t,y))}{\rho(t,y)}\frac{d}{dy}\frac{z'(x(t,y))}{\rho(t,y)}=
\]
\[
=\frac{z'(x(t,y))}{\rho(t,y)}\left(\frac{z''(x(t,y))x_{y}(t,y)}{\rho(t,y)}-\frac{z'(x(t,y))\rho_{y}(t,y)}{\rho^{2}(t,y)}\right)=
\]
\[
=\frac{z'(x(t,y))}{\rho(t,y)}\left(\frac{z''(x(t,y))}{z'(x(t,y))}-\frac{z'(x(t,y))\rho_{y}(t,y)}{\rho^{2}(t,y)}\right)=\frac{1}{\rho(t,y)}\left(z''(x(t,y))-\frac{[z'(x(t,y))]^{2}\rho_{y}(t,y)}{\rho^{2}(t,y)}\right).
\]
That is why the function $R(t,y)$ can be written in terms of the
density 
\[
R(t,y)=(\omega')^{2}\frac{1}{\rho(t,y)}\left(\frac{z''(x(t,y))-\frac{[z'(x(t,y))]^{2}\rho_{y}(t,y)}{\rho^{2}(t,y)}-z''(x(t,y))}{[z'(x(t,y))]^{2}}\right)=
\]
\[
=-(\omega')^{2}\frac{1}{\rho(t,y)}\frac{\rho_{y}(t,y)}{\rho^{2}(t,y)}=(\omega')^{2}\frac{1}{\rho(t,y)}\frac{d}{dy}\frac{1}{\rho(t,y)}
\]
Thus all assertions of the theorem are proved.

\subsection{Proof of the theorem on the force and energy}

\paragraph{Proof of the Theorem \ref{Th_force}}

Write down the force $R^{(N)}(t,y)$ in terms of $q$ variables, introduced
in the proof of Theorem \ref{Th_bound} 
\[
R^{(N)}(t,y)=\omega^{2}(q_{k(y,N,t)}-q_{k(y,N,t)-1}).
\]
By Lemma \ref{qklimit} we have: 
\[
R^{(N)}(t,y)=N(\omega')^{2}\left(q\left(t,\frac{k(y,N,t)}{N}\right)-q\left(t,\frac{k(y,N,t)-1}{N}\right)\right)+O(\frac{\ln N}{N})=
\]
\[
=(\omega')^{2}q_{x}\left(t,\frac{k(y,N,t)}{N}\right)+O(\frac{\ln N}{N})
\]
Using inequalities (\ref{k_N_bound_2}) and (\ref{k_N_bound_1}) we
have the following estimate: 
\[
\left|\frac{k(y,N,t)}{N}-z(x(t,y))\right|\leqslant\frac{c}{N}.
\]
for some constant $c$, not depending on $N$. Then we can conclude
that 
\[
\lim_{N\rightarrow\infty}R^{(N)}(t,y)=R(t,y)=(\omega')^{2}q_{x}(t,z(x(t,y)))=(\omega')^{2}\frac{d^{2}G(t,z(x(t,y)))}{dz^{2}}.
\]
Using formula (\ref{R_t_y}), we get the proof.

\paragraph{\.{P}roof of Theorem \ref{Th_Energy}}

Let us check the first equality. Rewrite the potential energy $U^{(N)}(t,y)$
in terms of the $q$ variables, which were introduced in the proof
of Theorem \ref{Th_bound} 
\[
U^{(N)}(t,y)=\frac{1}{4}\omega^{2}(q_{k_{N}(t,y)}^{2}+q_{k_{N}(t,y)-1}^{2}).
\]
The same arguments as in the proof of Theorem \ref{Th_force} give
\[
U(t,y)=\lim_{N\rightarrow\infty}U^{(N)}(t,y)=\frac{1}{2}(\omega')^{2}q^{2}(t,z(x(t,y)))=\frac{1}{2}(\omega')^{2}\left(\frac{dG(t,z(x(t,y)))}{dz}-1\right)^{2}.
\]
Using formulas (\ref{y_po_x}) and (\ref{yderivx}), we get: 
\[
U(t,y)=\frac{1}{2}(\omega')^{2}\left(\frac{y_{x}(x(t,y))}{z'(x(t,y))}-1\right)^{2}=\frac{1}{2}(\omega')^{2}\left(\frac{1}{\rho(t,y)}-1\right)^{2}
\]
The formula for the kinetic energy is obvious.

\section{The density dynamics}

On the three-dimensional $(t,x,z)\in R^{3}$ graph the surface $z=\rho(t,x)-1$
is presented, as the result of computer modelling with $N=200,\omega'=1$.
Initial data were chosen as: 
\[
X(x)=1+\epsilon S_{n}(x),\quad S_{n}(x)=\sum_{k=4}^{100}\frac{s_{k}}{k^{2}}\sin(\pi kx),V(x)=0
\]
with random numbers $s_{k}\in[0,1]$ . $\epsilon$ is chosen so that
there were no particle collisions, namely as 
\[
\epsilon<\frac{1}{2s},\quad s=\int_{0}^{1}|S''_{n}(x)|dx.
\]

\includegraphics[width=15cm,height=7cm]{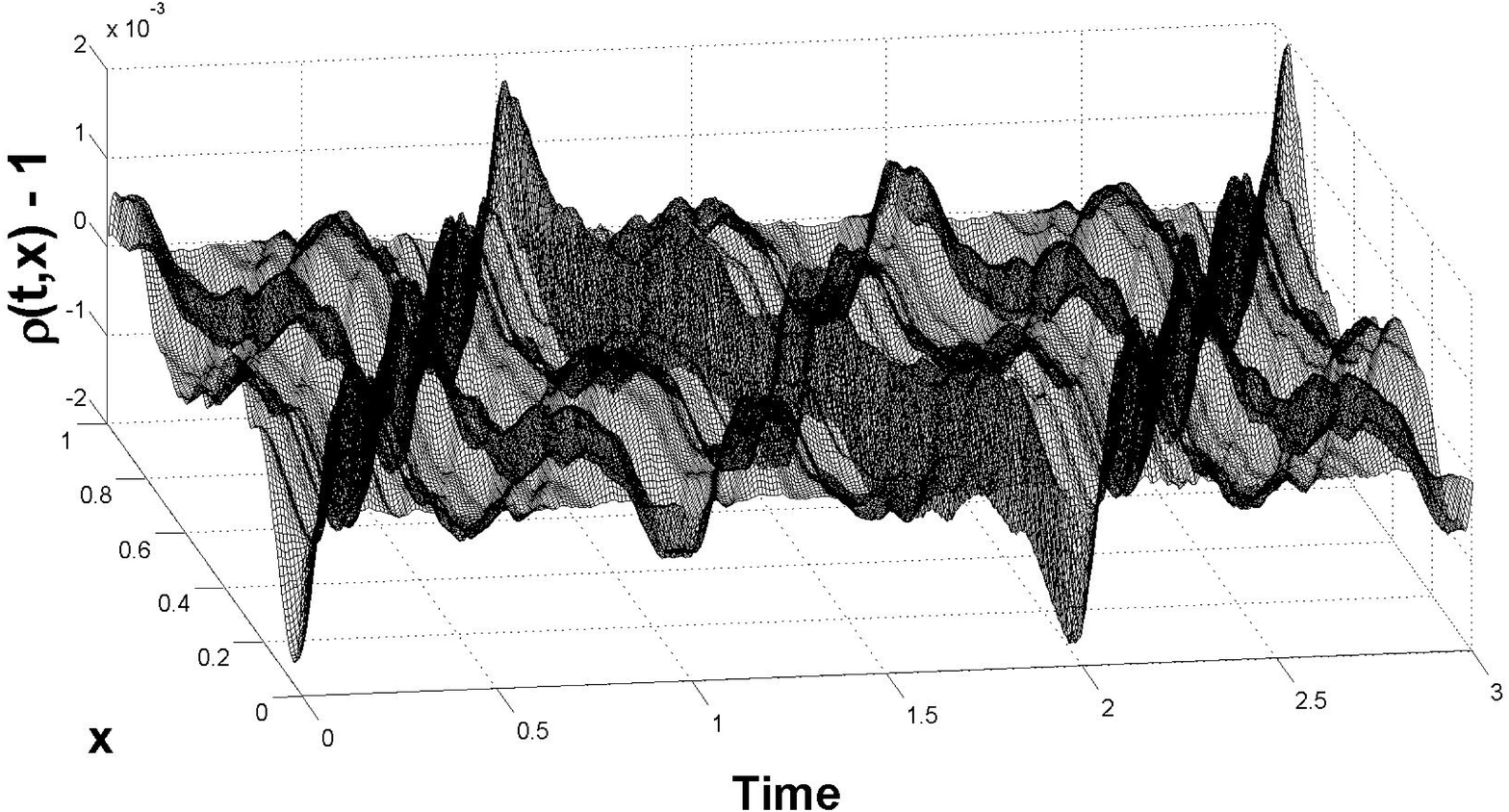}

\pagebreak


\begin{thebibliography}{1}
\bibitem{Bogoliubov} Bogolyubov N. N. On some statistical methods
in mathematical physics. 1845. Kiev. Acad of Science of USSR, 1945.

\bibitem{Morrey}Morrey C. On the derivation of the equations of hydrodynamics
from statistical mechanics. Comm. Pure Appl. Math., 1955, 8, 279-326.

\bibitem{March_Pulvirenti}Marchioro, M. Pulvirenti. Mathematical
Theory of Incompressible Nonviscous Fluids. Springer. 1993.

\bibitem{Esposito1}R. Esposito, J. Lebowitz, R. Marra. On the derivation
of hydrodynamics from the Boltzmann equation. Phys. of Fluids, 1999,
v. 11, 8, pp. 2354-2366.

\bibitem{Esposito2}R. Esposito, R. Marra. Incompressible fluids on
three levels: hydrodynamic, kinetic, microscopic. Mathematical Analysis
of Phenomena in Fluid and Plasma Dynamics , RIMS, Kyoto (1993).

\bibitem{Dobrushin}Boldrighini C., Dobrushin R. L., Sukhov Yu. M.
One-dimensional hard rod caricature of hydrodynamics. Journal of Statistical
Physics, 1983, 31, No, 3.

\bibitem{ChorinMarsden}Chorin A., Marsden J. A mathematical introduction
to fluid mechanics. Third Ed. Springer. 2000.

\bibitem{Siegel}Siegel C,. Moser J. Lectures on Celestial Mechanics.
Springer-Verlag. 1971. \end{thebibliography}
\end{document}